\documentclass[twocolumn,english]{IEEEtran}
\usepackage[T1]{fontenc}
\usepackage[latin9]{inputenc}
\usepackage{geometry}
\usepackage{subfigure}
\usepackage{amsmath}
\usepackage{graphicx}
\usepackage{amssymb}

\makeatletter
\newtheorem{prop}{Proposition}
\newtheorem{lemma}{Lemma}
\newtheorem{thm}{Theorem}
\newtheorem{remrk}{Remark}

\usepackage{cite}
\usepackage{url}
\usepackage{bm}
\usepackage{bbm}

\makeatother

\usepackage{babel}

\begin{document}

\title{How Many Users should be Turned On\\
in a Multi-Antenna Broadcast Channel?}

\author{Wei Dai, \emph{Member, IEEE}, Youjian (Eugene) Liu, \emph{Member,
IEEE}, Brian C. Rider and Wen Gao}

\maketitle
\begin{abstract}
This paper considers broadcast channels with $L$ antennas at the
base station and $m$ single-antenna users, where $L$ and $m$ are
typically of the same order. We assume that only partial channel state
information is available at the base station through a finite rate
feedback. Our key observation is that the optimal number of on-users
(users turned on), say $s$, is a function of signal-to-noise ratio
(SNR) and feedback rate. In support of this, an asymptotic analysis
is employed where $L$, $m$ and the feedback rate approach infinity
linearly. We derive the asymptotic optimal feedback strategy as well
as a realistic criterion to decide which users should be turned on.
The corresponding asymptotic throughput per antenna, which we define
as the \emph{spatial efficiency}, turns out to be a function of the
number of on-users $s$, and therefore $s$ must be chosen appropriately.
Based on the asymptotics, a scheme is developed for systems with finite
many antennas and users. Compared with other studies in which $s$
is presumed constant, our scheme achieves a significant gain. Furthermore,
our analysis and scheme are valid for heterogeneous systems where
different users may have different path loss coefficients and feedback
rates.
\end{abstract}
\renewcommand{\thefootnote}{\fnsymbol{footnote}} \footnotetext[0]{Manuscript received November 1, 2007; revised April 15, 2008. This work was supported by NSF Grants DMS-0505680, CCF-0728955, ECCS-0725915, and Thomson Inc. Part of content was presented at the Conference on Information Sciences and Systems,  Baltimore, MD, March 2007.

W. Dai is with the Department of Electrical and Computer Engineering, University of Illinois at Urbana-Champaign, Urbana, IL 61801, USA (email: wei.dai@colorado.edu).

Y. Liu and B. C. Rider are with the Department of Electrical and Computer Engineering and the Department of Mathematics, respectively, University of Colorado at Boulder, Boulder, CO 80309, USA (email: eugeneliu@ieee.org, brian.rider@colorado.edu).

W. Gao is with Thomson Inc., Two Independence Way, Princeton, NJ 08540, USA (email:wen.gao@thomson.net).} \renewcommand{\thefootnote}{\arabic{footnote}} \setcounter{footnote}{0}

\section{\label{sec:Introduction}Introduction}

It is well known that multiple antennas can improve the spectral efficiency.
This paper considers broadcast channels with $L$ antennas at the
base station and $m$ single-antenna users. To achieve the full benefit,
perfect channel state information (CSI) is required at both receiver
and transmitter. Perfect CSI at the receiver can be obtained by estimation
from the received signal. However, if CSI at the transmitter (CSIT)
is obtained from feedback, perfect CSIT requires an infinite feedback
rate. As this is not feasible in practice, it is important to analyze
the effect of finite rate feedback and design efficient strategy accordingly.

The feedback models for broadcast channels are described as follows.
To save feedback rate on power control, we assume a power on/off strategy%
\footnote{This assumption will be further validated in Section \ref{sec:System-Model}.%
} where each user is either turned on with a constant power or turned
off. For a given channel realization, the users quantize their channel
states into finite bits and feedback the corresponding indices to
the base station. After receiving the feedback from users, the base
station decides which users should be turned on and then forms beamforming
vectors for transmission. 

Broadcast channels with feedback have been widely studied recently.
Ideally, if the base station has the perfect CSI, dirty paper codes
or zero-forcing transmission can help clean off interference among
users. However, with only finite rate feedback on CSI, the base station
does not know the perfect channel state information and therefore
interference from other users is inevitable. The interference gets
so strong at high signal-to-noise ratio (SNR) regions that the system
throughput is upper bounded by a constant even when SNR approaches
infinity. This phenomena, called interference domination, was reported
on in \cite{Sharif_IT05_MIMO_BC_Feedback,Jindal_IT06_BC_Feedback}.
One way to combat it is to allow the number of users to be much larger
than the number of antennas at the base station. With sufficiently
many independent realizations of the channel, it is possible to obtain
$L$ orthogonal users with feedback: Sharif and Hassibi select users
whose channel directions are close to a random generated basis vectors
in \cite{Sharif_IT05_MIMO_BC_Feedback}; Yoo, et. al., pick up near
orthogonal users in an iterative way \cite{Jindal_ISIT2006_MIMO_BC_Feedback_Lots_Users,Jindal_JSAC2007_MultiAntenna_BC_Feedback_User_Selection}.
Recently, Bayesteh and Khandani quantified the feedback required as
a function of the number of users \cite{Khandani_ISIT2006_How_Much_Feedback_MIMO_BC,Khandani_ITsubmitted_How_Much_Feedback_MIMO_BC}.
Another approach is to fix both the number of antennas at the base
station and the system size (the number of users). It has been shown
in \cite{Shamai_Allerton2005_MIMO_BC_Imperfect_CSI} that the maximum
achievable multiplexing gain is one (at high SNR) with finite rate
feedback. The full multiplexing gain requires the feedback rate increases
linearly with SNR \cite{Jindal_IT06_BC_Feedback}. In both approaches,
a homogeneous system is assumed where all the users share the same
path loss coefficient and feedback resource. 

Separate from the above, this paper studies a more realistic scenario:

\begin{itemize}
\item We consider heterogeneous systems where different users may have different
path loss coefficients and feedback rates.
\item The size of the broadcast system is small. That is, the number of
users and the number of antennas at the base station are typically
of the same order. Note that a cooperative communication network can
often be viewed as a composition of multi-access and broadcast sub-systems
with a small number of users. Research on broadcast systems of small
size also provides insights into cooperative communications. 
\item Analysis and design are valid for arbitrary SNR. 
\end{itemize}
According to the authors' knowledge, the above practically important
scenario has not been systematically studied due to the associated
difficulty in analysis.

For such systems, we solve the interference domination problem by
choosing the appropriate number of on-users $s$. This solution comes
from an asymptotic analysis where $L,m,s$ and the feedback rates
approach infinity linearly. As have been demonstrated in \cite{Dai_05_Power_onoff_strategy_design_finite_rate_feedback}
and will be verified in our simulations in Fig. \ref{cap:Rate-ZF},
this type of asymptotic analysis is surprisingly reliable when being
applied to small systems. The main asymptotic results include: 

\begin{itemize}
\item It is asymptotically optimal to only quantize the channel directions
and ignore the channel magnitude information. The asymptotically optimal
feedback function and codebook are derived accordingly.
\item A realistic on/off criterion is proposed to decide which users should
be turned on.
\item The corresponding throughput per antenna converges to a constant,
defined as the \emph{spatial efficiency}. It is a function of the
normalized number of on-users $\bar{s}=\frac{s}{L}$. Further, there
exists a unique $\bar{s}\in\left(0,1\right)$ to maximize the the
spatial efficiency. 
\end{itemize}

Based on the insights obtained from the above asymptotic results,
we develop a scheme to choose the appropriate $s$ for systems with
finite $L$ and $m$. Simulations show that the gain achieved by choosing
$s$ is significant compared with the strategies where $s=L$ \cite{Jindal_IT06_BC_Feedback}.
In addition, our scheme has the following advantages.

\begin{itemize}
\item It is valid for heterogeneous systems. 
\item The associated computation complexity is low. In the proposed scheme,
the choice of on-users is independent of the channel realization,
and therefore there is no need to select on-users every fading block.
The computation complexity is much smaller than that of user selection
\cite{Khandani_ISIT2006_How_Much_Feedback_MIMO_BC,Khandani_ITsubmitted_How_Much_Feedback_MIMO_BC,Sharif_IT05_MIMO_BC_Feedback}.
\item Only on-users need to feedback CSI, which saves the precious feedback
resource. 
\end{itemize}

\section{\label{sec:System-Model}System Model}

Consider a broadcast channel with $L$ antennas at the base station
and $m$ single-antenna users. Assume that the base station employs
zero forcing transmitter. Let $\gamma_{i}\ge0$ ($1\le i\le m$) be
the path loss coefficient for user $i$. Then the received signal
$Y_{i}\in\mathbb{C}$ for user $i$ is given by \[
Y_{i}=\sqrt{\gamma_{i}}\mathbf{h}_{i}^{\dagger}\left(\sum_{j=1}^{m}\mathbf{q}_{j}X_{j}\right)+W_{i},\]
where $\mathbf{h}_{i}\in\mathbb{C}^{L\times1}$ is the channel state
vector for user, $\mathbf{q}_{j}\in\mathbb{C}^{L\times1}$ is the
zero-forcing beamforming vector for user $j$, $X_{j}\in\mathbb{C}$
is the source signal for the user $j$ and $W_{i}\in\mathbb{C}$ is
the circularly symmetric complex Gaussian noise with zero mean and
unit variance $\mathcal{CN}\left(0,1\right)$. Here, we assume that
$\mathbf{q}_{j}^{\dagger}\mathbf{q}_{j}=1$ and the Rayleigh block
fading channel model: the entries of $\mathbf{h}_{i}$ are independent
and identically distributed (i.i.d.) $\mathcal{CN}\left(0,1\right)$.
Without loss of generality, we assume that $L\le m$; if $L>m$, adding
$L-m$ users with $\gamma_{i}=0$ yields an equivalent system with
$L^{\prime}=m$. 

For the above broadcast system, it is natural to assume a total power
constraint \[
\sum_{i=1}^{m}\mathrm{E}\left[\left|X_{i}\right|^{2}\right]\leq\rho.\]
 Further, we assume a power on/off strategy with a constant number
of on-users as follows. 

\begin{description}
\item [{A1)}] Power on/off strategy: a source $X_{i}$ is either turned
on with a constant power $P_{\mathrm{on}}$ or turned off. 
\item [{A2)}] A constant number of on-users: we assume that the number
of on-users $s$ ($1\le s\le m$) is a constant independent of the
specific channel realizations, and thus $P_{\mathrm{on}}=\frac{\rho}{s}$.
Here, $s$ is allowed to be a function of SNR, which distinguishes
this paper from \cite{Sharif_IT05_MIMO_BC_Feedback,Jindal_IT06_BC_Feedback}
where $s=L$ always. 
\end{description}
A similar strategy has been demonstrated near optimal for single user
MIMO systems in our work \cite{Dai_05_Power_onoff_strategy_design_finite_rate_feedback}.
Although little is known about the optimality of the proposed strategy
in broadcast systems, we adopt it for two reasons: first, this strategy
has simple implementation and similar forms are employed in many practical
systems, see IEEE802.20 and IEEE802.22 for example; second, it saves
precious feedback resources on power control.

The finite rate feedback model is then described as follows. Assume
that both base station and user $i$ knows $\gamma_{i}$%
\footnote{There are many ways in which the base station obtains $\gamma_{i}$.
A simple example could be that the base station measures the feedback
signal strength. %
} but only user $i$ knows the channel state realization $\mathbf{h}_{i}$
perfectly. For given channel realizations $\mathbf{h}_{1}\cdots\mathbf{h}_{m}$,
an on-user $i$ quantizes his channel $\mathbf{h}_{i}$ into $R_{i}$
bits and then feeds the corresponding index to the base station. Formally,
let $\mathcal{B}_{i}=\left\{ \hat{\mathbf{h}}\in\mathbb{C}^{L\times1}\right\} $
with $\left|\mathcal{B}_{i}\right|=2^{R_{i}}$ be a channel state
codebook for user $i$. Then the quantization function is given by
$\mathfrak{q}\left(\mathbf{h}_{i},\mathcal{B}_{i}\right)=\hat{\mathbf{h}}_{i}$.
In Section \ref{sub:quantization-function} and \ref{sub:codebooks},
we will show how to design $\mathfrak{q}$ and $\mathcal{B}$ respectively.

After receiving feedback information from users, the base station
decides which $s$ users should be turned on and forms zero-forcing
beamforming vectors for them. Let $A_{\mathrm{on}}$ be the set of
the $s$ on-users. The zero-forcing beamforming vectors $\mathbf{q}_{i}$'s
$i\in A_{\mathrm{on}}$ is calculated as follows%
\footnote{Our interpretation of constructing zero forcing beamforming vectors
is different from the traditional one (see \cite{Jindal_JSAC2007_MultiAntenna_BC_Feedback_User_Selection}
for example). We adopt the unitary projection because not only does
it have an explicit geometric meaning but also it provides a nice
{}``isotropic'' property, which is crucial in proofs (see Appendix
\ref{sub:Signal-Energy-Calculation} and \ref{sub:Interference-Power-Calculation}
for details).%
}. Let $\mathcal{P}_{i}^{\perp}$ be the plane generated by \[
\left\{ \hat{\mathbf{h}}_{j}:\; j\in A_{\mathrm{on}}\backslash\left\{ i\right\} \right\} .\]
 Let $\mathcal{P}_{i}$ be the orthogonal complement of $\mathcal{P}_{i}^{\perp}$
and $t$ be the dimensions of $\mathcal{P}_{i}$. Let $\mathbf{T}_{i}\in\mathbb{C}^{L\times t}$
be a random matrix whose columns are orthonormal and span the plane
$\mathcal{P}$. Then $\mathbf{q}_{i}$ is the \emph{unitary projection}
of $\hat{\mathbf{h}}_{i}$on $\mathbf{T}_{i}$, that is, \[
\mathbf{q}_{i}:=\mathbf{T}_{i}\mathbf{T}_{i}^{\dagger}\hat{\mathbf{h}}_{i}/\left\Vert \mathbf{T}_{i}\mathbf{T}_{i}^{\dagger}\hat{\mathbf{h}}_{i}\right\Vert .\]
 Here, if $s=1$ and $A_{\mathrm{on}}=\left\{ i\right\} $, $\mathbf{T}_{i}$
is a $L\times L$ unitary matrix and \[
\mathbf{q}_{i}=\hat{\mathbf{h}}_{i}/\left\Vert \hat{\mathbf{h}}_{i}\right\Vert .\]

\section{\label{sec:Asymptotic-Analysis}Asymptotic Analysis }

As $m$ and $L$ are of the same order, we consider the asymptotic
region where $L,m,R_{i}\mathrm{'s}\rightarrow\infty$ linearly.

\subsection{\label{sub:quantization-function}Design of Quantization Function}

Generally speaking, full information of $\mathbf{h}_{i}$ contains
the direction information $\mathbf{v}_{i}:=\mathbf{h}_{i}/\left\Vert \mathbf{h}_{i}\right\Vert $
and the magnitude information $\left\Vert \mathbf{h}_{i}\right\Vert $.
In our Rayleigh fading channel model, it is well known that $\mathbf{v}_{i}$
and $\left\Vert \mathbf{h}_{i}\right\Vert $ are independent. Intuitively,
joint quantization of $\mathbf{v}_{i}$ and $\left\Vert \mathbf{h}_{i}\right\Vert $
is preferred.

Interestingly, Proposition \ref{pro:magnitude-concentration} implies
that there is no need to quantize the channel magnitudes. Indeed,
as $L,m\rightarrow\infty$ linearly, all users' channel magnitudes
concentrate on a single value in probability.

\begin{prop}
\label{pro:magnitude-concentration}For $\forall\epsilon>0$, as $L,m\rightarrow\infty$
with $\frac{m}{L}\rightarrow\bar{m}\in\mathbb{R}^{+}$, \[
\Pr\left(\underset{1\le i\le m}{\max}\;\frac{1}{L}\left\Vert \mathbf{h}_{i}\right\Vert ^{2}\ge1+\epsilon\right)\rightarrow0,\]
 and \[
\Pr\left(\underset{1\le i\le m}{\min}\;\frac{1}{L}\left\Vert \mathbf{h}_{i}\right\Vert ^{2}\le1-\epsilon\right)\rightarrow0.\]

\end{prop}

The proof is given in Appendix \ref{sub:proof-Thm-1}. It is noteworthy
that whether the users' channel magnitudes concentrate or not depends
on the relationship between $L$ and $m$: the concentration happens
when $L$ and $m$ are of the same order. To fully understand Proposition
\ref{pro:magnitude-concentration}, it is important to realize that
the Law of Large Numbers does not imply that all users' channel magnitudes
will concentrate uniformly. The Law of Large Numbers says that $\frac{1}{L}\left\Vert \mathbf{h}_{i}\right\Vert \rightarrow1$
almost surely for any \emph{given} $i$. However, if $m$ approaches
infinity exponentially with $L$, there are certain number of users
whose channel magnitudes are larger than others', and therefore it
may be still beneficial to quantize and feedback channel magnitude
information. Formally, consider a broadcast channel with $\gamma_{1}=\cdots=\gamma_{m}=1$.
As $L,m\rightarrow\infty$ with $\log\left(m\right)/L\rightarrow\bar{m}^{\prime}\in\mathbb{R}^{+}$,
there exists an $\epsilon>0,$ $\delta_{1}>0$ and $\delta_{2}>0$
such that \[
\frac{1}{L}\log\left|\left\{ i:\;\frac{1}{L}\left\Vert \mathbf{h}_{i}\right\Vert ^{2}>1+\frac{\epsilon}{2}\right\} \right|\rightarrow\delta_{1},\]
 and \[
\frac{1}{L}\log\left|\left\{ i:\;\frac{1}{L}\left\Vert \mathbf{h}_{i}\right\Vert ^{2}<1-\frac{\epsilon}{2}\right\} \right|\rightarrow\delta_{2}\]
 in probability. The proof follows from the standard large deviation
technique and is omitted here.

Proposition \ref{pro:magnitude-concentration} implies that it is
sufficient to quantize the channel direction information only and
omit the channel magnitude information. For this quantization, the
codebook is given by $\mathcal{B}_{i}=\left\{ \mathbf{p}\in\mathbb{C}^{L\times1}:\;\left\Vert \mathbf{p}\right\Vert =1\right\} $
with $\left|\mathcal{B}_{i}\right|=2^{R_{i}}$. Let $\mathbf{v}_{i}=\mathbf{h}_{i}/\left\Vert \mathbf{h}_{i}\right\Vert $.
The quantization output is given by\begin{equation}
\mathbf{p}_{i}=\mathfrak{q}\left(\mathbf{h}_{i},\mathcal{B}_{i}\right)=\underset{\mathbf{p}\in\mathcal{B}_{i}}{\arg\;\max}\;\left|\mathbf{v}_{i}^{\dagger}\mathbf{p}\right|.\label{eq:quantization-fn}\end{equation}

\subsection{\label{sub:codebooks}Asymptotically Optimal Codebooks}

Consider design of codebooks. Given the quantization function (\ref{eq:quantization-fn}),
the distortion of a given codebook $\mathcal{B}_{i}$ is the average
chordal distance between the actual and quantized channel directions
corresponding to the codebook $\mathcal{B}_{i}$ and defined as \[
D\left(\mathcal{B}_{i}\right):=1-\mathrm{E}_{\mathbf{h}_{i}}\left[\underset{\mathbf{p}\in\mathcal{B}_{i}}{\max}\left|\mathbf{v}_{i}^{\dagger}\mathbf{p}\right|^{2}\right].\]
 The following lemma bounds the minimum achievable distortion for
a given codebook rate. 

\begin{lemma}
\label{lem:dist-rate-fn}Define $D^{*}\left(R\right)\triangleq\underset{\mathcal{B}:\;\left|\mathcal{B}\right|\leq2^{R}}{\inf}D\left(\mathcal{B}\right)$.
Then\begin{align}
 & \frac{L-1}{L}2^{-\frac{R}{L-1}}\left(1+o\left(1\right)\right)\leq D^{*}\left(R\right)\nonumber \\
 & \qquad\leq\frac{\Gamma\left(\frac{1}{L-1}\right)}{L-1}2^{-\frac{R}{L-1}}\left(1+o\left(1\right)\right),\label{eq:quantization-bds}\end{align}
and as $L$ and $R$ approach infinity with $\frac{R}{L}\rightarrow\bar{r}\in\mathbb{R}^{+}$,
\[
\underset{\left(L,R\right)\rightarrow\infty}{\lim}D^{*}\left(R\right)=2^{-\bar{r}}.\]

\end{lemma}

The following Lemma shows that a random codebook is asymptotically
optimal in probability.

\begin{lemma}
\label{lem:random-codes-asymptotically-optimal}Let $\mathcal{B}_{\mathrm{rand}}$
be a random codebook where the vectors $\mathbf{p}\in\mathcal{B}_{\mathrm{rand}}$'s
are independently generated from the isotropic distribution. Let $R=\log\left|\mathcal{B}_{\mathrm{rand}}\right|$.
As $L,R\rightarrow\infty$ with $\frac{R}{L}\rightarrow\bar{r}\in\mathbb{R}^{+}$,
for $\forall\epsilon>0$, \[
\underset{\left(L,R\right)\rightarrow\infty}{\lim}\;\mathrm{Pr}\left\{ \mathcal{B}_{\mathrm{rand}}:\; D\left(\mathcal{B}_{\mathrm{rand}}\right)>2^{-\bar{r}}+\epsilon\right\} =0.\]

\end{lemma}

The proofs of Lemma \ref{lem:dist-rate-fn} and \ref{lem:random-codes-asymptotically-optimal}
are given in our paper \cite{Dai_IT2008_Quantization_Grassmannian_manifold}.
Due to the asymptotic optimality of random codebooks, we assume that
the codebooks $\mathcal{B}_{i}$'s $i=1,\cdots,m$ are independent
and randomly constructed throughout this paper.

\subsection{\label{sub:On/off-Criterion}On/off Criterion}

After receiving feedback from users, the base station should decide
which $s$ users should be turned on. 

Ideally, for given channel realizations $\mathbf{h}_{1},\cdots,\mathbf{h}_{m}$,
the optimal set of on users $A_{\mathrm{on}}^{*}$ should be chosen
to maximize the instantaneous mutual information. However, finding
$A_{\mathrm{on}}^{*}$ requires exhaustive search, whose complexity
exponentially increases with $m$. 

A suboptimal option is the random orthonormal beams construction method
in \cite{Sharif_IT05_MIMO_BC_Feedback}: the base station randomly
constructs $L$ orthonormal beams $\mathbf{b}_{1},\cdots,\mathbf{b}_{L}$,
finds the users with highest signal-to-noise-plus-interference ratios
(SINRs) through feedback from users, and then transmits to these selected
users. Note that the maximum SINR achievable for user $i$ is $\underset{1\le k\le L}{\max}\left|\mathbf{h}_{i}^{\dagger}\mathbf{b}_{k}\right|$.
However, Proposition \ref{pro:Random-Beams} below shows that in our
asymptotic region where $L,m\rightarrow\infty$ linearly, all users'
channels are near orthogonal to all of the $L$ orthonormal beams
$\mathbf{b}_{i}$'s. Therefore, all users' maximum SINRs approach
zero uniformly in probability, and no user should be turned on in
probability. The method in \cite{Sharif_IT05_MIMO_BC_Feedback} fails
in our asymptotic region.

\begin{prop}
\label{pro:Random-Beams}Given $\forall\epsilon>0$ and any $L$ orthonormal
beams $\mathbf{b}_{k}\in\mathbb{C}^{L\times1}$ $1\le k\le L$, as
$L,m\rightarrow\infty$ linearly with $\frac{m}{L}\rightarrow\bar{m}\in\mathbb{R}^{+}$,
\[
\underset{\left(L,m\right)\rightarrow\infty}{\lim}\;\Pr\left(\underset{1\le i\le m,\;1\le k\le L}{\max}\;\frac{1}{L}\left|\mathbf{h}_{i}^{\dagger}\mathbf{b}_{k}\right|>\epsilon\right)=0.\]

\end{prop}
\begin{proof}
See Appendix \ref{sub:Proof-of-Thm-Random-Beams}. 
\end{proof}

In this paper, we take another approach where the on/off decision
is independent of channel directions. We start with the throughput
analysis for a specific on-user $i\in A_{\mathrm{on}}$. Note that
\[
Y_{i}=\sqrt{\gamma_{i}}\mathbf{h}_{i}^{\dagger}\mathbf{q}_{i}X_{i}+\left(\sqrt{\gamma_{i}}\mathbf{h}_{i}^{\dagger}\sum_{j\in A_{\mathrm{on}}\backslash\left\{ i\right\} }\mathbf{q}_{j}X_{j}+W\right).\]
The signal power and interference power for user $i$ are given by
\begin{equation}
P_{\mathrm{sig},i}=\frac{\rho}{s}\gamma_{i}\left|\mathbf{h}_{i}^{\dagger}\mathbf{q}_{i}\right|^{2}\label{eq:signal-power}\end{equation}
 and \begin{equation}
P_{\mathrm{int},i}=\frac{\rho}{s}\gamma_{i}\sum_{j\in A_{\mathrm{on}}\backslash\left\{ i\right\} }\left|\mathbf{h}_{i}^{\dagger}\mathbf{q}_{j}\right|^{2}\label{eq:interference-power}\end{equation}
respectively. If the choice of $A_{\mathrm{on}}$ is independent of
the channel directions $\mathbf{v}_{i}$'s, we have a nice property
regarding to $P_{\mathrm{sig},i}$ and $P_{\mathrm{int},i}$. 

\begin{thm}
\label{thm:on-user-i-throughput}Let $\left|A_{\mathrm{on}}\right|=s$
be chosen independently of $\mathbf{v}_{i}$'s. Let $L,m,s,R_{i}\mathrm{'s}\rightarrow\infty$
with $\frac{m}{L}\rightarrow\bar{m}\in\mathbb{R}^{+}$, $\frac{s}{L}\rightarrow\bar{s}\in\left[0,1\right]$
and $\frac{R_{i}}{L}\rightarrow\bar{r}_{i}\in\mathbb{R}^{+}$. Assume
that $\mathbf{v}_{i}$'s $i\in A_{\mathrm{on}}$ are independent.
Then for $\forall i\in A_{\mathrm{on}}$, \[
P_{\mathrm{sig},i}\rightarrow\frac{\rho}{\bar{s}}\gamma_{i}\left(1-2^{-\bar{r}_{i}}\right)\left(1-\bar{s}\right),\]
\[
P_{\mathrm{int},i}\rightarrow\rho\gamma_{i}2^{-\bar{r}_{i}},\]
 and therefore the throughput of user $i$ satisfies\[
\mathcal{I}_{i}:=\log\left(1+\frac{P_{\mathrm{sig},i}}{1+P_{\mathrm{int},i}}\right)\rightarrow\log\left(1+\eta_{i}\frac{1-\bar{s}}{\bar{s}}\right),\]
in probability, where\begin{equation}
\eta_{i}:=\frac{\rho\gamma_{i}\left(1-2^{-\bar{r}_{i}}\right)}{1+\rho\gamma_{i}2^{-\bar{r}_{i}}}.\label{eq:eta-zf}\end{equation}

\end{thm}
\begin{proof}
See Appendix \ref{sub:Signal-Energy-Calculation} and \ref{sub:Interference-Power-Calculation}. 
\end{proof}

Theorem \ref{thm:on-user-i-throughput} shows that if $A_{\mathrm{on}}$
is independent of $\mathbf{v}_{i}$'s, $\mathcal{I}_{i}$ is a function
of $\eta_{i}$ but independent of the specific channel realization
$\mathbf{h}_{i}$ in probability. Based on this fact, we select the
set of $s$ on-users $A_{\mathrm{on}}$ such that $\left|A_{\mathrm{on}}\right|=s$
and \begin{equation}
A_{\mathrm{on}}=\left\{ i:\;\eta_{i}\ge\eta_{j}\;\mathrm{for}\;\forall j\notin A_{\mathrm{on}}\right\} ;\label{eq:on-off-criterion}\end{equation}
if there are multiple candidates, we randomly choose one of them.
It is the asymptotically optimal on/off selection if the on/off decision
is independent of the channel direction information. The difference
between the throughput achieved by optimal on/off criterion (requiring
exhaustive search) and the proposed (\ref{eq:on-off-criterion}) remains
unknown.

\subsection{\label{sub:The-Spatial-Efficiency}The Spatial Efficiency}

We define \emph{the spatial efficiency (bits/sec/Hz/antenna)} as \[
\bar{\mathcal{I}}\left(\bar{s}\right):=\underset{\left(L,m,s,R_{i}\mathrm{'s}\right)\rightarrow\infty}{\lim}\bar{\mathcal{I}}^{\left(L\right)},\]
 where $L,m,s,R_{i}\mathrm{'s}\rightarrow\infty$ in the same way
as before, $\bar{\mathcal{I}}^{\left(L\right)}$ is the average throughput
per antenna given by \[
\bar{\mathcal{I}}^{\left(L\right)}:=\mathrm{E}_{\mathcal{B}_{i}\mathrm{'s},\mathbf{h}_{i}\mathrm{'s}}\left[\frac{1}{L}\sum_{i\in A_{\mathrm{on}}}\log\left(1+\frac{P_{\mathrm{sig},i}}{1+P_{\mathrm{int},i}}\right)\right],\]
and $A_{\mathrm{on}}$, $P_{\mathrm{sig},i}$ and $P_{\mathrm{int},i}$
are defined in (\ref{eq:on-off-criterion}), (\ref{eq:signal-power})
and (\ref{eq:interference-power}) respectively. 

We shall quantify $\bar{\mathcal{I}}\left(\bar{s}\right)$ for a given
$\bar{s}$. Define the empirical distribution of $\eta_{i}$ as \[
\mu_{\eta}^{\left(m\right)}\left(\eta\le x\right):=\frac{1}{m}\left|\left\{ \eta_{i}:\;\eta_{i}\le x\right\} \right|,\]
 and assume that $\mu_{\eta}:=\lim\mu_{\eta}^{\left(m\right)}$ exists
weakly as $L,m,R_{i}\mathrm{'s}\rightarrow\infty$. In order to cope
with $\mu_{\eta}$'s with mass points, define \[
\int_{x^{+}}^{\infty}f\left(\eta\right)d\mu_{\eta}:=\underset{\Delta x\downarrow0}{\lim}\int_{x+\Delta x}^{\infty}f\left(\eta\right)d\mu_{\eta}\]
for $\forall x\in\mathbb{R}$, where $f$ is a integrable function
with respect to $\mu_{\eta}$. Then $\bar{\mathcal{I}}\left(\bar{s}\right)$
is computed in the following theorem.

\begin{thm}
\label{thm:spatial-efficiency}Let $L,m,s,R_{i}\mathrm{'s}\rightarrow\infty$
with $\frac{m}{L}\rightarrow\bar{m}$, $\frac{s}{L}\rightarrow\bar{s}$
and $\frac{R_{i}}{L}\rightarrow\bar{r}_{i}$. Define \[
\eta_{\bar{s}}:=\sup\left\{ \eta:\;\bar{m}\int_{\eta}^{\infty}d\mu_{\eta}>\bar{s}\right\} .\]
 Then as $\bar{s}\notin\left(0,1\right)$, $\bar{\mathcal{I}}\left(\bar{s}\right)=0$.
If $\bar{s}\in\left(0,1\right)$,\begin{align}
\bar{\mathcal{I}}\left(\bar{s}\right) & =\bar{m}\int_{\eta_{\bar{s}}^{+}}^{\infty}\log\left(1+\eta\frac{1-\bar{s}}{\bar{s}}\right)d\mu_{\eta}\nonumber \\
 & +\left(\bar{s}-\bar{m}\int_{\eta_{\bar{s}}^{+}}^{\infty}d\mu_{\eta}\right)\log\left(1+\eta_{\bar{s}}\frac{1-\bar{s}}{\bar{s}}\right).\label{eq:zf-spatial-efficiency}\end{align}

\end{thm}
\begin{proof}
It actually follows from Theorem \ref{thm:on-user-i-throughput}.
\end{proof}

We are also interested in finding the optimal $\bar{s}$ to maximize
$\bar{\mathcal{I}}\left(\bar{s}\right)$. Though $\bar{\mathcal{I}}\left(\bar{s}\right)$
is not a concave function in general, the following theorem provides
a criterion to find the optimal $\bar{s}$. 

\begin{thm}
\label{thm:optimal-s}$\bar{\mathcal{I}}\left(\bar{s}\right)$ is
maximized at a unique $\bar{s}^{*}\in\left(0,1\right)$ such that\begin{align}
0\in & \left[\underset{\Delta\bar{s}\rightarrow0}{\lim\;\inf}\frac{\bar{\mathcal{I}}\left(\bar{s}^{*}\right)-\bar{\mathcal{I}}\left(\bar{s}^{*}-\Delta\bar{s}\right)}{\Delta\bar{s}},\right.\nonumber \\
 & \left.\underset{\Delta\bar{s}\rightarrow0}{\lim\;\sup}\frac{\bar{\mathcal{I}}\left(\bar{s}^{*}\right)-\bar{\mathcal{I}}\left(\bar{s}^{*}-\Delta\bar{s}\right)}{\Delta\bar{s}}\right].\label{eq:optimal-s}\end{align}

\end{thm}

The proof is in Appendix \ref{sub:Proof-of-Optimal-s}. The corresponding
$\bar{\mathcal{I}}\left(\bar{s}^{*}\right)$ is the maximum achievable
spatial efficiency for the proposed power on/off strategy. It is noteworthy
that $\bar{s}^{*}$ is not a monotone function of SNR $\rho$ according
to our empirical calculation.

\section{\label{sec:Finite-System-Design}Finite Dimensional System Design}

Based on the above asymptotic results, we now propose a scheme for
systems with finite $L$ and $m$.

\subsection{Throughput Estimation for Finite Dimensional Systems}

While asymptotic analysis provide many insights, we do not apply asymptotic
results directly for a finite dimensional system. The reason is that
in asymptotic analysis $\frac{1}{L}\rightarrow0$ while $\frac{1}{L}>0$
for finite dimensional systems. To see the difference more explicitly,
let us calculate the main order term of the throughput for user $i\in A_{\mathrm{on}}$.
For user $i\in A_{\mathrm{on}}$, the corresponding throughput is
\begin{align*}
\mathcal{I}_{i} & =\mathrm{E}\left[\log\left(1+\frac{P_{\mathrm{sig},i}}{1+P_{\mathrm{int},i}}\right)\right]\\
 & =\log\left(1+\frac{\mathrm{E}\left[P_{\mathrm{sig},i}\right]}{1+\mathrm{E}\left[P_{\mathrm{int},i}\right]}\right)\\
 & \quad+\mathrm{E}\left[\log\left(\frac{1+P_{\mathrm{sig},i}+P_{\mathrm{int},i}}{1+\mathrm{E}\left[P_{\mathrm{sig},i}\right]+\mathrm{E}\left[P_{\mathrm{int},i}\right]}\right)\right]\\
 & \quad-\mathrm{E}\left[\log\left(\frac{1+P_{\mathrm{int},i}}{1+\mathrm{E}\left[P_{\mathrm{int},i}\right]}\right)\right],\end{align*}
where $P_{\mathrm{sig},i}$ and $P_{\mathrm{int},i}$ are defined
in (\ref{eq:signal-power}) and (\ref{eq:interference-power}). We
quantify $\mathrm{E}\left[P_{\mathrm{sig},i}\right]$ and $\mathrm{E}\left[P_{\mathrm{int},i}\right]$
in below.

\begin{thm}
\label{thm:Average-sig-int-power}Let $\mathcal{B}_{i}$'s be randomly
constructed and $D_{i}=\mathrm{E}_{\mathcal{B}_{i}}\left[D\left(\mathcal{B}_{i}\right)\right]$
for all $1\le i\le m$. For randomly chosen $A_{\mathrm{on}}$ and
$i\in A_{\mathrm{on}}$, if $1\le s\le L$\begin{align}
\mathrm{E}\left[P_{\mathrm{sig},i}\right] & =\gamma_{i}\rho\frac{L}{s}\left[\left(1-D_{i}\right)\left(1-\frac{s-1}{L}\right)\right.\nonumber \\
 & \quad\left.+D_{i}\frac{s-1}{L\left(L-1\right)}\right],\label{eq:zf-average-signal}\end{align}
 and \begin{equation}
\mathrm{E}\left[P_{\mathrm{int},i}\right]=\gamma_{i}\rho\frac{L}{s}\frac{s-1}{L-1}D_{i};\label{eq:zf-average-interference}\end{equation}
if $s>L$, $\mathrm{E}\left[P_{\mathrm{sig},i}\right]=0$. 
\end{thm}

The proof is provided in Appendix \ref{sub:Signal-Energy-Calculation}
and \ref{sub:Interference-Power-Calculation}. Define 

\begin{equation}
\mathcal{I}_{\mathrm{main},i}:=\log\left(1+\frac{\mathrm{E}\left[P_{\mathrm{sig},i}\right]}{1+\mathrm{E}\left[P_{\mathrm{int},i}\right]}\right).\label{eq:main-order-term}\end{equation}
It can be verified from Theorem \ref{thm:on-user-i-throughput} that
$\mathcal{I}_{i}=\mathcal{I}_{\mathrm{main},i}+o\left(1\right)$ and
therefore $\mathcal{I}_{\mathrm{main},i}$ is the main order term
of $\mathcal{I}_{i}$. Then the difference between asymptotic analysis
and finite dimensional systems analysis is clear. In the limit, $\frac{s-1}{L}\rightarrow\bar{s}$
and $\frac{R_{i}}{L-1}\rightarrow\bar{r}_{i}$. However, for finite
dimensional systems, simply substituting the asymptotic values into
(\ref{eq:zf-average-signal}-\ref{eq:main-order-term}) directly introduces
unpleasant error, especially when $L$ is small.  Therefore, to estimate
$\mathcal{I}_{i}$ ($\forall i\in A_{\mathrm{on}}$) for finite dimensional
systems, we have to rely on (\ref{eq:zf-average-signal})-(\ref{eq:main-order-term}).

The calculation of $\mathrm{E}\left[P_{\mathrm{sig},i}\right]$ and
$\mathrm{E}\left[P_{\mathrm{int},i}\right]$ relies on quantification
of $D_{i}$. In general, it is difficult to compute $D_{i}$ precisely.
Note that the upper bound in (\ref{eq:quantization-bds}) is derived
by evaluating the average performance of random codebooks (see \cite{Dai_IT2008_Quantization_Grassmannian_manifold}
for details). We use its main order term to estimate $D_{i}$: \[
D_{i}\approx\frac{\Gamma\left(\frac{1}{L-1}\right)}{L-1}2^{-\frac{R_{i}}{L-1}}.\]

\subsection{\label{sub:Scheme-Finite-Case}A Scheme for Finite Dimensional Systems}

Given system parameters, a practical scheme finding the appropriate
$s$ and $A_{\mathrm{on}}$ is developed. 

For a given $s$, the set of $A_{\mathrm{on}}$ is decided as follows:
calculate $\mathcal{I}_{\mathrm{main},1},\cdots,\mathcal{I}_{\mathrm{main},m}$
according to (\ref{eq:main-order-term}) and choose the $s$ users
with the largest $\mathcal{I}_{\mathrm{main},i}$'s to turn on; if
there exists an ambiguity, random selection is employed to resolve
it. For example, if $\mathcal{I}_{\mathrm{main},1}=\mathcal{I}_{\mathrm{main},2}=\cdots=\mathcal{I}_{\mathrm{main},m}$,
the $s$ on-users are randomly drawn from all the $m$ users. Note
again, that $A_{\mathrm{on}}$ is independent of the channel realization.

The appropriate $s$ is chosen as follows. Let \[
\mathcal{I}_{\mathrm{main}}\left(s\right)=\underset{A_{\mathrm{on}}:\;\left|A_{\mathrm{on}}\right|=s}{\max}\;\sum_{i\in A_{\mathrm{on}}}\mathcal{I}_{\mathrm{main},i}.\]
We choose the number of on-users to be \[
s_{\mathrm{main}}^{*}=\underset{1\le s\le L}{\arg\;\max}\;\mathcal{I}_{\mathrm{main}}\left(s\right).\]

Although the above procedure involves exhaustive search, the corresponding
complexity is actually low. First, the calculations are independent
of instantaneous channel realizations. Only system parameters $L$,
$m$, $\gamma_{i}$'s, $R_{i}$'s and $\rho$, are needed. Provided
that $\gamma_{i}$'s change slowly, the base station does not need
to recalculate $s_{\mathrm{main}}^{*}$ and $A_{\mathrm{on}}$ frequently.
Second, $R_{i}=R_{j}$ in most systems. For such systems, the $s$
on-users are just simply the users with the largest $\gamma_{i}$'s. 

After calculating $s_{\mathrm{main}}^{*}$ and $A_{\mathrm{on}}$,
the base station broadcast $A_{\mathrm{on}}$ to all the users. For
each fading block, the system works as follows. 1) At the beginning
of each fading block, the base station broadcasts a single channel
training sequence to help all the users estimate their channel states
$\mathbf{h}_{i}$'s. 2) After estimating their $\mathbf{h}_{i}$'s,
the on-users quantize $\mathbf{h}_{i}$'s into $\mathbf{p}_{i}$'s
according to (\ref{eq:quantization-fn}) and feed the corresponding
indices to the base station. 3) The base station then calculates the
transmit beamforming vectors $\mathbf{q}_{i}$'s and transmits $\mathbf{q}_{i}X_{i}$'s. 

\begin{remrk}
[Fairness Scheduling]For systems with $\gamma_{i}\neq\gamma_{j}$
or $R_{i}\neq R_{j}$, there may be some users always turned off according
to the above scheme. Fairness scheduling is therefore needed to ensure
fairness of the system. An example could be as follows. Given $m$
users, calculate the corresponding $s_{\mathrm{main}}^{*}$ and $A_{\mathrm{on}}$,
and then turns on the users in $A_{\mathrm{on}}$ for the first fading
block. At the second fading block, only consider the users who have
not been turned on $\left\{ 1,\cdots,m\right\} \backslash A_{\mathrm{on}}$.
Calculate the corresponding $s_{\mathrm{main}}^{*}$ and $A_{\mathrm{on}}$,
and then turns on the users in the new $A_{\mathrm{on}}$. Proceed
this process until all users have been turned on once. Then start
a new scheduling cycle.
\end{remrk}

\subsection{Simulation Results}

\begin{figure}
\subfigure[$R_{\mathrm{fb}}=6$ Bits/Channel Realization]{\includegraphics[scale=0.5]{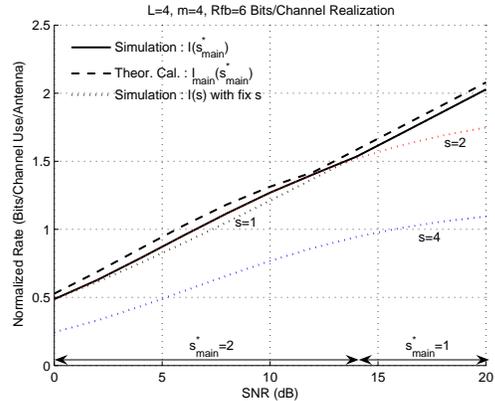}}

\subfigure[$R_{\mathrm{fb}}=12$ Bits/Channel Realization]{\includegraphics[scale=0.5]{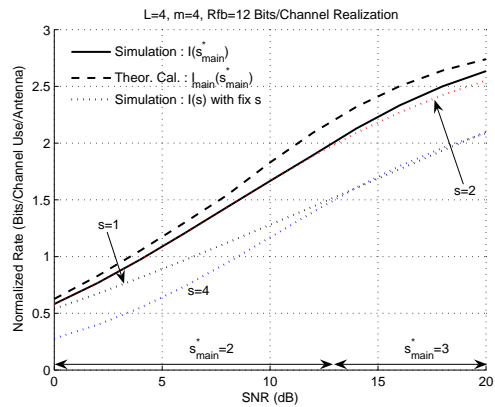}}

\caption{\label{cap:Rate-ZF}Total Throughput for Zero Forcing Beamforming}

\end{figure}

Fig. \ref{cap:Rate-ZF} gives the simulation results for the proposed
scheme using zero-forcing. In the simulations, $L=m=4$. For simplicity,
we assume that $\gamma_{1}=\gamma_{2}=\cdots=\gamma_{m}=1$ and $R_{1}=R_{2}=\cdots=R_{m}=R_{\mathrm{fb}}$.
With these assumptions, the $s$ on-users can be randomly chosen from
all the $m$ users. Without loss of generality, we assume that $A_{\mathrm{on}}\equiv\left\{ 1,\cdots,s\right\} $.
Let $\mathcal{I}\left(s\right)=\sum_{i\in A_{\mathrm{on}}}\mathcal{I}_{i}.$
In Fig. \ref{cap:Rate-ZF}, the solid lines are the simulations of
$\mathcal{I}\left(s_{\mathrm{main}}^{*}\right)$ while the dashed
lines are the theoretical calculation of $\mathcal{I}_{\mathrm{main}}\left(s_{\mathrm{main}}^{*}\right)$.
The simulation results show that the optimal $s$ is a function of
$\rho$ and $R_{\mathrm{fb}}$. For example, $s=1$ is optimal when
$\rho\in\left[15,20\right]$dB and $R_{\mathrm{fb}}=6$ bits, while
$s=3$ is optimal for the same SNR region as $R_{\mathrm{fb}}$ increases
to 12 bits. The reason behind it is that the interference introduced
by finite rate quantization is larger when $R_{\mathrm{fb}}$ is smaller:
when $R_{\mathrm{fb}}$ is small, the base station needs to turn off
some users to avoid strong interference as SNR gets very large. 

We also compare our scheme with the schemes where the number of on-users
is a presumed constant (independent of $\rho$ and $R_{\mathrm{fb}}$).
The throughput of schemes with presumed $s$ is presented in dotted
lines. From the simulation results, the throughput achieved by choosing
appropriate $s$ is always better than or equals to that with presumed
$s$. Specifically, compared to the scheme in \cite{Jindal_IT06_BC_Feedback}
where $s=L=4$ always, our scheme achieves a significant gain at high
SNR by turning off some users.

It is interesting to observe that given feedback rates, the optimal
number of on-users $s_{\mathrm{main}}^{*}$ is \emph{not} monotonic
with SNR $\rho$. As discussed before, $s_{\mathrm{main}}^{*}=1$
as $\rho\rightarrow\infty$ to avoid the interference domination phenomenon.
While $\rho\rightarrow0$, it can be shown that $s_{\mathrm{main}}^{*}=1$
as well. For this case, compared with the noise power, the interference
is weak and can be ignorable. Setting $s=1$ avoids signal power loss
(the $-\frac{s-1}{L}$ term in (\ref{eq:zf-average-signal})) due
to zero forcing projection and therefore is optimal. For median SNRs,
we have to rely on the scheme in Section \ref{sub:Scheme-Finite-Case}.

\section{\label{sec:Conclusion}Conclusion}

This paper considers heterogeneous broadcast systems with a relatively
small number of users. Asymptotic analysis where $L,m,s,R_{i}\rightarrow\infty$
linearly is employed to get insight into system design. We derive
the asymptotically optimal feedback strategy, propose a realistic
on/off criterion, and quantify the spatial efficiency. The key observation
is that the number of on-users should be appropriately chosen as a
function of system parameters. Finally, a practical scheme is developed
for finite dimensional systems. Simulations show that this scheme
achieves a significant gain compared with previously studied schemes
with presumed number of on-users. 

\appendix

\subsection{\label{sub:proof-Thm-1}Proof of Proposition \ref{pro:magnitude-concentration}}

This proposition is proved by standard large deviation argument. Note
that $\left\Vert \mathbf{h}_{i}\right\Vert $'s $\left(i=1,\cdots,m\right)$
are independent and identically distributed.\begin{align*}
 & \Pr\left(\underset{1\le i\le m}{\max}\frac{1}{L}\left\Vert \mathbf{h}_{i}\right\Vert ^{2}\ge1+\epsilon\right)\\
 & =1-\left(\Pr\left(\frac{1}{L}\left\Vert \mathbf{h}_{1}\right\Vert ^{2}<1+\epsilon\right)\right)^{m}\\
 & =1-\exp\left\{ m\log\left(1-\Pr\left(\frac{1}{L}\left\Vert \mathbf{h}_{1}\right\Vert ^{2}\ge1+\epsilon\right)\right)\right\} .\end{align*}
 For all $\alpha\in\left(0,1\right)$, by Chebyshev's inequality,
\begin{align*}
 & \Pr\left(\frac{1}{L}\left\Vert \mathbf{h}_{1}\right\Vert ^{2}\ge1+\epsilon\right)\\
 & =\Pr\left(\sum_{l=1}^{L}\left(\left|h_{1,l}\right|^{2}-1\right)\ge L\epsilon\right)\\
 & \le\exp\left\{ -L\left(\alpha\epsilon-\log\mathrm{E}\left[e^{\alpha\left(\left|h_{1,1}\right|^{2}-1\right)}\right]\right)\right\} \\
 & =\exp\left\{ -L\left(\alpha\left(1+\epsilon\right)+\log\left(1-\alpha\right)\right)\right\} .\end{align*}
Take $\alpha=\frac{\epsilon}{1+\epsilon}$. We have \[
\Pr\left(\frac{1}{L}\left\Vert \mathbf{h}_{1}\right\Vert ^{2}\ge1+\epsilon\right)\le\exp\left\{ -L\left(\epsilon-\log\left(1+\epsilon\right)\right)\right\} .\]
 Let $f^{+}\left(\epsilon\right):=\epsilon-\log\left(1+\epsilon\right)$.
It can be verified that $f^{+}\left(\epsilon\right)>0$ for $\epsilon>0$.
Thus, for any given $\delta>0$, if $L$ is sufficiently large, \begin{align*}
 & \Pr\left(\underset{1\le i\le m}{\max}\frac{1}{L}\left\Vert \mathbf{h}_{i}\right\Vert ^{2}\ge1+\epsilon\right)\\
 & \le1-\exp\left(\bar{m}L\left(1+o\left(1\right)\right)\log\left(1-\exp\left(-Lf^{+}\left(\epsilon\right)\right)\right)\right)\\
 & =1-\exp\left(-\bar{m}Le^{-Lf^{+}\left(\epsilon\right)}\left(1+o\left(1\right)\right)\right)\le\delta,\end{align*}
which proves the first part of Proposition \ref{pro:magnitude-concentration}. 

The second part is proved similarly. For any given $\delta>0$, \begin{align*}
 & \Pr\left(\underset{1\le i\le m}{\min}\frac{1}{L}\left\Vert \mathbf{h}_{i}\right\Vert ^{2}\le1-\epsilon\right)\\
 & =1-\exp\left\{ m\log\left(1-\Pr\left(\frac{1}{L}\left\Vert \mathbf{h}_{i}\right\Vert ^{2}\le1-\epsilon\right)\right)\right\} \\
 & \overset{\left(a\right)}{\le}1-\exp\left\{ m\log\left(1-e^{-L\left(\alpha\left(1-\epsilon\right)+\log\left(1-\epsilon\right)\right)}\right)\right\} \\
 & \overset{\left(b\right)}{=}1-\exp\left\{ m\log\left(1-e^{-L\left(-\log\left(1-\epsilon\right)-\epsilon\right)}\right)\right\} \\
 & \overset{\left(c\right)}{=}1-\exp\left\{ -\bar{m}Le^{-L\left(-\log\left(1-\epsilon\right)-\epsilon\right)}\left(1+o\left(1\right)\right)\right\} \le\delta,\end{align*}
 where $\left(a\right)$ holds for all $\alpha\in\left(-1,0\right)$
(by Chebyshev's inequality), $\left(b\right)$ follows from setting
$\alpha=-\frac{\epsilon}{1-\epsilon}$, and $\left(c\right)$ follows
from the fact that $-\log\left(1-\epsilon\right)-\epsilon>0$ for
$\epsilon\in\left(0,1\right)$ and the Taylor's expansion of $\log\left(1-x\right)$.

\subsection{\label{sub:Proof-of-Thm-Random-Beams}Proof of Proposition \ref{pro:Random-Beams}}

This proposition is based on the observation that $\left|\mathbf{h}_{i}^{\dagger}\mathbf{b}_{k}\right|\sim\mathcal{CN}\left(0,1\right)$
are i.i.d. $\left(1\le i\le m,\;1\le k\le L\right)$. Let $\mathbf{B}=\left[\mathbf{b}_{1}\cdots\mathbf{b}_{L}\right]$.
Then the above observation is verified by $\mathrm{E}\left[\left(\mathbf{B}^{\dagger}\mathbf{h}_{i}\right)\left(\mathbf{B}^{\dagger}\mathbf{h}_{i}\right)\right]=\mathbf{I},$
and $\mathrm{E}\left[\left(\mathbf{B}^{\dagger}\mathbf{h}_{i}\right)\left(\mathbf{B}^{\dagger}\mathbf{h}_{j}\right)\right]=\mathbf{0}$
for $i\ne j$. Note that \[
\Pr\left(\left|\mathbf{h}_{1}^{\dagger}\mathbf{b}_{1}\right|^{2}>L\epsilon\right)=e^{-L\epsilon}.\]
 For any given $\delta>0$, as $L$ is sufficiently large, \begin{align*}
 & \Pr\left(\underset{1\le i\le m,1\le k\le L}{\max}\;\frac{1}{L}\left|\mathbf{h}_{i}^{\dagger}\mathbf{b}_{k}\right|^{2}>\epsilon\right)\\
 & =1-\left(1-\Pr\left(\left|\mathbf{h}_{1}^{\dagger}\mathbf{b}_{1}\right|^{2}>L\epsilon\right)\right)^{mL}\\
 & =1-\exp\left\{ \bar{m}L^{2}\left(1+o\left(1\right)\right)\phantom{\left(\left|\mathbf{h}_{1}^{\dagger}\right|^{2}\right)}\right.\\
 & \quad\quad\left.\cdot\log\left(1-\Pr\left(\left|\mathbf{h}_{1}^{\dagger}\mathbf{b}_{1}\right|^{2}>L\epsilon\right)\right)\right\} \\
 & =1-\exp\left\{ -\bar{m}L^{2}e^{-L\epsilon}\left(1+o\left(1\right)\right)\right\} \le\delta,\end{align*}
which completes the proof.

\subsection{\label{sub:Signal-Energy-Calculation}Signal Energy Calculation}

The signal power can be written as\begin{align*}
\frac{1}{L}\left|\mathbf{h}_{1}^{\dagger}\mathbf{q}_{1}\right|^{2} & =\frac{1}{L}\left|\mathbf{h}_{1}^{\dagger}\left[\mathbf{p}_{1}\mathbf{p}_{1}^{\perp}\right]\left[\mathbf{p}_{1}\mathbf{p}_{1}^{\perp}\right]^{\dagger}\mathbf{q}_{1}\right|^{2}\\
 & =\frac{1}{L}\left|\mathbf{h}_{1}^{\dagger}\mathbf{p}_{1}\mathbf{p}_{1}^{\dagger}\mathbf{q}_{1}\right|^{2}+\frac{1}{L}\left|\mathbf{h}_{1}^{\dagger}\mathbf{p}_{1}^{\perp}\left(\mathbf{p}_{1}^{\perp}\right)^{\dagger}\mathbf{q}_{1}\right|^{2}\\
 & \quad+\frac{1}{L}\left(\mathbf{h}_{1}^{\dagger}\mathbf{p}_{1}\mathbf{p}_{1}^{\dagger}\mathbf{q}_{1}\right)\left(\mathbf{q}_{1}^{\dagger}\mathbf{p}_{1}^{\perp}\left(\mathbf{p}_{1}^{\perp}\right)^{\dagger}\mathbf{h}_{1}\right)\\
 & \quad+\frac{1}{L}\left(\mathbf{h}_{1}^{\dagger}\mathbf{p}_{1}^{\perp}\left(\mathbf{p}_{1}^{\perp}\right)^{\dagger}\mathbf{q}_{1}\right)\left(\mathbf{q}_{1}^{\dagger}\mathbf{p}_{1}\mathbf{p}_{1}^{\dagger}\mathbf{h}_{1}\right).\end{align*}

\subsubsection{Asymptotic Analysis}

Here, we prove that $\frac{1}{L}\mathbf{h}_{1}^{\dagger}\mathbf{q}_{1}\mathbf{q}_{1}^{\dagger}\mathbf{h}_{1}\rightarrow\left(1-\bar{s}\right)\left(1-2^{\bar{r}_{1}}\right)$.
It is an application of the following Lemma \ref{lem:Signal-Power-Asymp-01}-\ref{lem:Signal-Power-Asymp-03}. 

\begin{lemma}
\label{lem:Signal-Power-Asymp-01}$\frac{1}{L}\left|\mathbf{h}_{1}^{\dagger}\mathbf{p}_{1}\mathbf{p}_{1}^{\dagger}\mathbf{q}_{1}\right|^{2}\rightarrow\left(1-\bar{s}\right)\left(1-2^{-\bar{r}}\right).$
\end{lemma}
\begin{proof}
We claim that \[
\frac{1}{L}\left|\mathbf{h}_{1}^{\dagger}\mathbf{p}_{1}\right|^{2}\rightarrow1-2^{-\bar{r}}\]
 in probability. It is follows from the facts that $\frac{1}{L}\left\Vert \mathbf{h}\right\Vert ^{2}\rightarrow1$
in probability and that $\mathbf{v}_{1}^{\dagger}\mathbf{p}_{1}\mathbf{p}_{1}^{\dagger}\mathbf{v}_{1}\rightarrow1-2^{-\bar{r}}$
(Lemma \ref{lem:random-codes-asymptotically-optimal}). We shall show
that \[
\left|\mathbf{p}_{1}^{\dagger}\mathbf{q}_{1}\right|^{2}\rightarrow1-\bar{s}\]
 in probability. Note that $\mathbf{p}_{1}$ and $\mathbf{T}_{1}$
are isotropically distributed and independent. The statistics of $\mathbf{T}_{1}^{\dagger}\mathbf{p}_{1}$
is the same as that of \[
\frac{1}{\left\Vert \mathbf{h}\right\Vert /\sqrt{L}}\frac{1}{\sqrt{L}}\mathbf{T}_{1}^{\dagger}\mathbf{h}^{\prime}\]
 where $\mathbf{h}^{\prime}\in\mathbb{C}^{L\times1}$ is a random
Gaussian vector with independent $\mathcal{CN}\left(0,1\right)$ entries.
Note that $\mathbf{T}_{1}$ has rank $L-\left(s-1\right)$ with probability
one. $\mathbf{T}_{1}^{\dagger}\mathbf{h}^{\prime}$ contains $L-s+1$
i.i.d. $\mathcal{CN}\left(0,1\right)$ entries with probability one.
It follows that $\frac{1}{L}\left\Vert \mathbf{h}^{\prime}\right\Vert ^{2}\rightarrow1$,
\[
\frac{1}{L}\left\Vert \mathbf{T}_{1}^{\dagger}\mathbf{h}^{\prime}\right\Vert ^{2}\rightarrow1-\bar{s}\]
 and \[
\left\Vert \mathbf{T}_{1}^{\dagger}\mathbf{p}_{1}\right\Vert ^{2}\rightarrow1-\bar{s}\]
 in probability. Hence, \[
\left|\mathbf{p}_{1}^{\dagger}\mathbf{q}_{1}\right|^{2}=\mathbf{p}_{1}^{\dagger}\mathbf{T}_{1}\mathbf{T}_{1}^{\dagger}\mathbf{p}_{1}\rightarrow1-\bar{s}\]
 in probability.
\end{proof}

\begin{lemma}
\label{lem:Signal-Power-Asymp-02}\[
\frac{1}{L}\left|\mathbf{h}_{1}^{\dagger}\mathbf{p}_{1}^{\perp}\left(\mathbf{p}_{1}^{\perp}\right)^{\dagger}\mathbf{q}_{1}\right|^{2}\rightarrow0\]
 in probability.
\end{lemma}
\begin{proof}
Suppose that $\mathbf{p}_{1}$ is given. Without loss of generality,
assume that $\mathbf{p}_{1}=\left[1,0,\cdots,0\right]^{\dagger}$%
\footnote{If $\mathbf{p}_{1}$ does not have the claimed form, we then apply
the rotation $\left[\mathbf{p}_{1}\mathbf{p}_{1}^{\perp}\right]^{\dagger}$
for some $\mathbf{p}_{1}^{\perp}$ to $\mathbf{h}_{1},\cdots,\mathbf{h}_{s}$
and $\mathcal{B}_{1},\cdots,\mathcal{B}_{s}$. This rotation gives
$\mathbf{p}_{1}^{\prime}=\mathfrak{q}\left(\mathbf{h}_{1}^{\prime},\mathcal{B}_{1}^{\prime}\right)=\left[1,0,\cdots,0\right]^{\dagger}$
but will not change the analysis.%
}. Let \[
\mathbf{w}:=\mathbf{p}_{1}^{\perp}\left(\mathbf{p}_{1}^{\perp}\right)^{\dagger}\mathbf{h}_{1}/\left\Vert \mathbf{p}_{1}^{\perp}\left(\mathbf{p}_{1}^{\perp}\right)^{\dagger}\mathbf{h}_{1}\right\Vert \]
 be the unitary projection of $\mathbf{h}_{1}$ on $\mathbf{p}_{1}^{\perp}$.
Then $\mathbf{w}$ has the form $\left[0,w_{1},\cdots,w_{L-1}\right]^{\dagger}$.
We shall show it is invariantly distributed under the rotation \[
\mathcal{U}_{L}^{1}:=\left[\begin{array}{cc}
1\\
 & \mathcal{U}_{L-1}\end{array}\right]\]
 as follows. Let \[
\mathcal{H}_{1}:=\left\{ \mathbf{h}_{1}:\;\mathfrak{q}\left(\mathbf{h}_{1},\mathcal{B}_{1}\right)=\mathbf{p}_{1}\right\} .\]
 Note that for any $\mathbf{U}_{L}^{1}\in\mathcal{U}_{L}^{1}$, \[
\mathfrak{q}\left(\mathbf{U}_{L}^{1}\mathbf{h}_{1},\mathbf{U}_{L}^{1}\mathcal{B}\right)=\mathbf{U}_{L}^{1}\mathbf{p}_{1}=\mathbf{p}_{1}.\]
 $\mathcal{H}_{1}$ is invariantly distributed under $\mathcal{U}_{L}^{1}$.
Further, $\mathbf{p}_{1}^{\perp}$ is also invariantly distributed
under $\mathcal{U}_{L}^{1}$. Since $\mathbf{w}$ is nothing but the
unitary projection of $\mathbf{h}_{1}$ on $\mathbf{p}_{1}^{\perp}$,
$\mathbf{w}$ is invariantly distributed under $\mathcal{U}_{L}^{1}$
(see also \cite{James_54_Normal_Multivariate_Analysis_Orthogonal_Group})
. Hence, the statistics of $\mathbf{w}$ is the same as that of \[
\frac{\sqrt{L-1}}{\left\Vert \mathbf{h}_{L-1}^{\prime}\right\Vert }\frac{1}{\sqrt{L-1}}\left[\begin{array}{c}
0\\
\mathbf{h}_{L-1}^{\prime}\end{array}\right]\]
 where $\mathbf{h}_{L-1}^{\prime}\in\mathbb{C}^{\left(L-1\right)\times1}$
is a random standard Gaussian vector. It can be verified that for
any given $\mathbf{q}\in\mathbb{C}^{L\times1}$ with unit norm, \[
\frac{1}{\sqrt{L-1}}\left[0,\mathbf{h}_{L-1}^{\prime\phantom{L-}^{\dagger}}\right]\mathbf{q}\rightarrow0\]
 in probability. Now note that \[
\frac{1}{L}\left\Vert \mathbf{p}_{1}^{\perp}\left(\mathbf{p}_{1}^{\perp}\right)^{\dagger}\mathbf{h}_{1}\right\Vert ^{2}\rightarrow2^{-\bar{r}}\]
 and \[
\tfrac{\left\Vert \mathbf{h}_{L-1}^{\prime}\right\Vert }{\sqrt{L-1}}\rightarrow1\]
 in probability. This Lemma is proved. 
\end{proof}

\begin{lemma}
\label{lem:Signal-Power-Asymp-03}\[
\frac{1}{L}\left(\mathbf{h}_{1}^{\dagger}\mathbf{p}_{1}\mathbf{p}_{1}^{\dagger}\mathbf{q}_{1}\right)\left(\mathbf{q}_{1}^{\dagger}\mathbf{p}_{1}^{\perp}\left(\mathbf{p}_{1}^{\perp}\right)^{\dagger}\mathbf{h}_{1}\right)\rightarrow0\]
 in probability.
\end{lemma}
\begin{proof}
It follows from that \[
\frac{1}{L}\left|\mathbf{h}_{1}^{\dagger}\mathbf{p}_{1}\mathbf{p}_{1}^{\dagger}\mathbf{q}_{1}\right|^{2}\rightarrow c<\infty\]
 and \[
\frac{1}{L}\left|\mathbf{q}_{1}^{\dagger}\mathbf{p}_{1}^{\perp}\left(\mathbf{p}_{1}^{\perp}\right)^{\dagger}\mathbf{h}_{1}\right|^{2}\rightarrow0\]
 in probability.
\end{proof}

\subsubsection{Finite Dimensional Analysis}

For finite dimensional system, we shall show that \[
\mathrm{E}\left[\frac{1}{L}\mathbf{h}_{1}^{\dagger}\mathbf{q}_{1}\mathbf{q}_{1}^{\dagger}\mathbf{h}_{1}\right]=D_{1}\frac{s-1}{L\left(L-1\right)}.\]
This result is proved by combining Lemma \ref{lem:Signal-Power-Finite-01}-\ref{lem:Signal-Power-Finite-04}.

\begin{lemma}
\label{lem:Signal-Power-Finite-01}Given $\mathbf{p}_{1}$, \[
\mathrm{E}_{\mathbf{T}_{1}}\left[\left|\mathbf{p}_{1}^{\dagger}\mathbf{q}_{1}\right|^{2}\right]=\frac{L-s+1}{L}.\]
 Furthermore, \[
\mathrm{E}_{\mathbf{h}_{1},\mathcal{B}_{1}}\left[\frac{1}{L}\left|\mathbf{h}_{1}^{\dagger}\mathbf{p}_{1}\right|^{\dagger}\right]=1-D_{1}.\]

\end{lemma}
\begin{proof}
Given $\mathbf{p}_{1}$, \[
\left|\mathbf{p}_{1}^{\dagger}\mathbf{q}_{1}\right|^{2}=\mathbf{p}_{1}^{\dagger}\mathbf{T}_{1}\mathbf{T}_{1}^{\dagger}\mathbf{p}_{1}.\]
Note that $\mathbf{T}_{1}\in\mathcal{U}_{L\times\left(L-s+1\right)}$
with probability one, is isotropically distributed and independent
of $\mathbf{p}_{1}$. By arguments on the Grassmann manifold \cite{Dai_IT2008_Quantization_Grassmannian_manifold},
it can be verified that \[
\mathrm{E}_{\mathbf{T}_{1}}\left[\left|\mathbf{p}_{1}^{\dagger}\mathbf{q}_{1}\right|^{2}\right]=\frac{L-s+1}{L}.\]
 
\end{proof}

\begin{lemma}
\label{lem:Signal-Power-Finite-02}Given $\mathbf{p}_{1}$ and $\mathbf{p}_{1}^{\perp}$,
\[
\mathrm{E}_{\mathbf{T}_{1}}\left[\left(\mathbf{p}_{1}^{\perp}\right)^{\dagger}\mathbf{q}_{1}\mathbf{q}_{1}^{\dagger}\mathbf{p}_{1}^{\perp}\right]=\frac{s-1}{L\left(L-1\right)}\mathbf{I}_{L-1}.\]
 
\end{lemma}
\begin{proof}
For any given $\mathbf{V}\in\mathcal{U}_{L-1}$, let \begin{equation}
\mathbf{U}=\left[\mathbf{p}\mathbf{p}^{\perp}\right]\left[\begin{array}{cc}
1\\
 & \mathbf{V}\end{array}\right]\left[\mathbf{p}\mathbf{p}^{\perp}\right]^{\dagger}.\label{eq:V_maps_to_U}\end{equation}
Then $\mathbf{U}\in\mathcal{U}_{L}$, $\mathbf{U}\mathbf{p}^{\perp}=\mathbf{p}^{\perp}\mathbf{V}$
and $\mathbf{U}\mathbf{p}=\mathbf{p}$. Let $\mathbf{T}\in\mathcal{U}_{L\times\left(L-s+1\right)}$
be isotropically distributed and independent of $\mathbf{p}$ and
$\mathbf{p}^{\perp}$. Then\begin{align}
 & \mathrm{E}_{\mathbf{T}}\left[\left(\mathbf{p}^{\perp}\right)^{\dagger}\frac{\mathbf{T}\mathbf{T}^{\dagger}\mathbf{p}}{\left\Vert \mathbf{T}\mathbf{T}^{\dagger}\mathbf{p}\right\Vert }\left(\frac{\mathbf{T}\mathbf{T}^{\dagger}\mathbf{p}}{\left\Vert \mathbf{T}\mathbf{T}^{\dagger}\mathbf{p}\right\Vert }\right)^{\dagger}\mathbf{p}^{\perp}\right]\nonumber \\
 & \overset{\left(a\right)}{=}\mathrm{E}_{\mathbf{U}\mathbf{T}}\left[\left(\mathbf{p}^{\perp}\right)^{\dagger}\frac{\mathbf{T}\mathbf{T}^{\dagger}\mathbf{p}}{\left\Vert \mathbf{T}\mathbf{T}^{\dagger}\mathbf{p}\right\Vert }\cdots\right]\nonumber \\
 & =\mathrm{E}_{\mathbf{U}\mathbf{T}}\left[\left(\mathbf{U}\mathbf{p}^{\perp}\right)^{\dagger}\frac{\mathbf{U}\mathbf{T}\left(\mathbf{U}\mathbf{T}\right)^{\dagger}\mathbf{U}\mathbf{p}}{\left\Vert \mathbf{U}\mathbf{T}\left(\mathbf{U}\mathbf{T}\right)^{\dagger}\mathbf{U}\mathbf{p}\right\Vert }\cdots\right]\nonumber \\
 & \overset{\left(b\right)}{=}\mathrm{E}_{\mathbf{T}}\left[\left(\mathbf{U}\mathbf{p}^{\perp}\right)^{\dagger}\frac{\mathbf{T}\mathbf{T}^{\dagger}\mathbf{U}\mathbf{p}}{\left\Vert \mathbf{T}\mathbf{T}^{\dagger}\mathbf{U}\mathbf{p}\right\Vert }\cdots\right]\nonumber \\
 & =\mathbf{V}^{\dagger}\mathrm{E}_{\mathbf{T}}\left[\left(\mathbf{p}^{\perp}\right)^{\dagger}\frac{\mathbf{T}\mathbf{T}^{\dagger}\mathbf{p}}{\left\Vert \mathbf{T}\mathbf{T}^{\dagger}\mathbf{p}\right\Vert }\cdots\right]\mathbf{V},\label{eq:invariant-V-signal-power}\end{align}
where $\left(a\right)$ follows from the fact that $\mathbf{T}$ is
isotropically distributed and therefore $d\mu_{\mathbf{T}}=d\mu_{\mathbf{UT}}$,
and $\left(b\right)$ follows from the variable change from $\mathbf{UT}$
to $\mathbf{T}$. Since (\ref{eq:invariant-V-signal-power}) is valid
for arbitrary $\mathbf{V}\in\mathcal{U}_{L-1}$, $\mathrm{E}_{\mathbf{T}}\left[\cdots\right]=c\mathbf{I}$
for some constant $c\ge0$. 

We calculate $c$ as follows. Note that $\mathbf{q}^{\dagger}\left[\mathbf{p}\mathbf{p}^{\perp}\right]\left[\mathbf{p}\mathbf{p}^{\perp}\right]^{\dagger}\mathbf{q}=1$.
Then \begin{align*}
c & =\frac{1}{L-1}\mathrm{E}\left[\mathrm{tr}\left(\left(\mathbf{p}^{\perp}\right)^{\dagger}\mathbf{q}\mathbf{q}^{\dagger}\mathbf{p}^{\perp}\right)\right]\\
 & =\frac{1}{L-1}\mathrm{E}\left[\mathbf{q}^{\dagger}\mathbf{p}^{\perp}\left(\mathbf{p}^{\perp}\right)^{\dagger}\mathbf{q}\right]\\
 & =\frac{1-\frac{L-s+1}{L}}{L-1}=\frac{s-1}{L\left(L-1\right)}.\end{align*}

\end{proof}
\begin{lemma}
\label{lem:Signal-Power-Finite-03}Given $\mathbf{p}_{1}$ and $\mathbf{p}_{1}^{\perp}$,
\[
\mathrm{E}_{\mathbf{h}_{1},\mathcal{B}_{1}}\left[\left(\mathbf{p}_{1}^{\perp}\right)^{\dagger}\mathbf{h}_{1}\mathbf{h}_{1}^{\dagger}\mathbf{p}_{1}^{\perp}\right]=\frac{D}{L-1}\mathbf{I}_{L-1}.\]

\end{lemma}
\begin{proof}
For an arbitrary $\mathbf{V}\in\mathcal{U}_{L-1}$, let $\mathbf{U}\in\mathcal{U}_{L}$
be in (\ref{eq:V_maps_to_U}). Then \[
\mathfrak{q}\left(\mathbf{U}\mathbf{h}_{1},\mathbf{U}\mathcal{B}_{1}\right)=\mathbf{U}\mathfrak{q}\left(\mathbf{h}_{1},\mathcal{B}_{1}\right)=\mathbf{U}\mathbf{p}_{1}=\mathbf{p}_{1}.\]
 By following the same idea of the proof of Lemma \ref{lem:Signal-Power-Finite-02},
this lemma is proved.
\end{proof}
\begin{lemma}
\label{lem:Signal-Power-Finite-04}Given $\mathbf{p}_{1}$ and $\mathbf{p}_{1}^{\perp}$,
\[
\mathrm{E_{\mathbf{T}_{1}}}\left[\mathbf{p}_{1}^{\dagger}\mathbf{q}_{1}\mathbf{q}_{1}^{\dagger}\mathbf{p}_{1}^{\perp}\right]=\mathbf{0}^{\dagger}\]
 and \[
\mathrm{E}_{\mathbf{h}_{1},\mathcal{B}_{1}}\left[\mathbf{p}_{1}^{\dagger}\mathbf{h}_{1}\mathbf{h}_{1}^{\dagger}\mathbf{p}_{1}^{\perp}\right]=\mathbf{0}^{\dagger}.\]
 
\end{lemma}
\begin{proof}
By the same method in Lemma \ref{lem:Signal-Power-Finite-02} and
\ref{lem:Signal-Power-Finite-03}, for an arbitrary $\mathbf{V}\in\mathcal{U}_{L-1}$,
$\mathrm{E}\left[\cdots\right]=\mathrm{E}\left[\cdots\right]\mathbf{V}$,
which holds if and only if $\mathrm{E}\left[\cdots\right]=\mathbf{0}^{\dagger}$. 
\end{proof}

\subsection{\label{sub:Interference-Power-Calculation}Interference Power Calculation}

The interference from user $j$ to user $1$ can be written as The
signal power can be written as

\begin{align*}
\frac{1}{L}\left|\mathbf{h}_{1}^{\dagger}\mathbf{q}_{j}\right|^{2} & =\frac{1}{L}\left|\mathbf{h}_{1}^{\dagger}\left[\mathbf{p}_{1}\mathbf{p}_{1}^{\perp}\right]\left[\mathbf{p}_{1}\mathbf{p}_{1}^{\perp}\right]^{\dagger}\mathbf{q}_{j}\right|^{2}\\
 & =\frac{1}{L}\left|\mathbf{h}_{1}^{\dagger}\mathbf{p}_{1}^{\perp}\left(\mathbf{p}_{1}^{\perp}\right)^{\dagger}\mathbf{q}_{j}\right|^{2},\end{align*}
where the last step follows from the construction $\mathbf{q}_{j}\perp\mathbf{p}_{1}$.
The total interference at user 1 is then \[
\frac{1}{L}\sum_{j=2}^{s}\left|\mathbf{h}_{1}^{\dagger}\mathbf{p}_{1}^{\perp}\left(\mathbf{p}_{1}^{\perp}\right)^{\dagger}\mathbf{q}_{j}\right|^{2}.\]

\subsubsection{Asymptotic Analysis}

Without loss of generality, assume that $\mathbf{p}=\left[1,0,\cdots,0\right]^{\dagger}$.
We have analyzed the property of $\mathbf{h}_{1}^{\dagger}\mathbf{p}_{1}^{\perp}\left(\mathbf{p}_{1}^{\perp}\right)^{\dagger}$
in the proof of Lemma \ref{lem:Signal-Power-Asymp-02}. It has been
shown there that the statistics of $\frac{1}{\sqrt{L}}\mathbf{h}_{1}^{\dagger}\mathbf{p}_{1}^{\perp}\left(\mathbf{p}_{1}^{\perp}\right)^{\dagger}$
is the same as that of $X_{L}\frac{1}{\sqrt{L-1}}\mathbf{h}_{L-1}^{\prime}$,
where \[
X_{L}=\left\Vert \frac{1}{\sqrt{L}}\mathbf{h}_{1}^{\dagger}\mathbf{p}_{1}^{\perp}\left(\mathbf{p}_{1}^{\perp}\right)^{\dagger}\right\Vert \frac{\sqrt{L-1}}{\left\Vert \mathbf{h}_{L-1}^{\prime}\right\Vert }\rightarrow\sqrt{2^{-\bar{r}}}\]
 in probability and $\mathbf{h}_{L-1}^{\prime}\in\mathbb{C}^{\left(L-1\right)\times1}$
is a standard Gaussian vector. Now for any given $2\le j\le s$, since
$\mathbf{q}_{j}\perp\mathbf{p}_{1}$ and $\left\Vert \mathbf{q}_{j}\right\Vert =1$,
the statistics of $\mathbf{q}_{j}$ is the same as that of \[
\frac{\sqrt{L-1}}{\left\Vert \mathbf{h}_{L-1}^{\prime\prime}\right\Vert }\frac{1}{\sqrt{L-1}}\left[0,\mathbf{h}_{L-1}^{\prime\prime\phantom{\;}^{\dagger}}\right]^{\dagger},\]
 where $\mathbf{h}_{L-1}^{\prime\prime}\in\mathbb{C}^{\left(L-1\right)\times1}$
is another standard Gaussian vector. $\mathbf{h}_{L-1}^{\prime}\mathbf{q}_{j}\sim\mathcal{CN}\left(0,1\right)$.
It then can be verified that \[
\frac{1}{L-1}\sum_{j=2}^{s}\left|\mathbf{h}_{L-1}^{\prime}\mathbf{q}_{j}\right|^{2}\rightarrow\bar{s}\]
 in probability. Therefore, the total interference converges to $\bar{s}2^{-\bar{r}}$
in probability.

\subsubsection{Finite Dimensional Analysis}

It can be shown that the average interference power is\begin{align*}
 & \frac{1}{L}\mathrm{tr}\left(\mathrm{E}\left[\left(\left(\mathbf{p}_{1}^{\perp}\right)^{\dagger}\mathbf{h}_{1}\mathbf{h}_{1}^{\dagger}\mathbf{p}_{1}^{\perp}\right)\right.\right.\\
 & \quad\left.\left.\cdot\left(\left(\mathbf{p}_{1}^{\perp}\right)^{\dagger}\sum_{j=2}^{s}\left(\mathbf{q}_{j}\mathbf{q}_{j}^{\dagger}\right)\mathbf{p}_{1}^{\perp}\right)\right]\right).\end{align*}
Given $\mathbf{p}_{1}$ and $\mathbf{p}_{1}^{\perp}$, it can be shown
that \[
\sum_{j=2}^{s-1}\mathrm{E}_{\mathbf{h}_{j}'s,\mathcal{B}_{j}'s}\left[\left(\mathbf{p}_{1}^{\perp}\right)^{\dagger}\mathbf{q}_{j}\mathbf{q}_{j}^{\dagger}\mathbf{p}_{1}^{\perp}\right]=\frac{s-1}{L-1}\mathbf{I}_{L-1}\]
 by the same technique in the proof of Lemma \ref{lem:Signal-Power-Finite-02}.
Combining this fact and Lemma \ref{lem:Signal-Power-Finite-03}, calculates
the average interference power.

\subsection{\label{sub:Proof-of-Optimal-s}Proof of Theorem \ref{thm:optimal-s}}

Here, we only prove Theorem \ref{thm:optimal-s} by assuming that
$d\mu_{\eta}$ contains no mass point. The proof for $d\mu_{\eta}$
containing mass points follows the same line but is much more complicated
and omitted due to the space limitation. For compositional convenience,
we use the following notations: $f\left(\eta,s\right)=\log\left(1+\eta\frac{1-s}{s}\right)$,
$f^{\prime}\left(\eta,s\right)=\frac{\partial f\left(\eta,s\right)}{\partial s}$,
and $y\left(s\right)=\int_{t}^{+\infty}f\left(\eta,s\right)d\mu_{\eta}$
where $t$ is given by $\inf\left\{ t:\;\int_{t}^{\infty}d\mu_{\eta}<s\right\} $.
When $d\mu_{\eta}$ contains no mass point, (\ref{eq:optimal-s})
is reduced to $0=y^{\prime}\left(s\right)$. To proceed, we need Lemma
\ref{lem:derivative}-\ref{lem:formula-1og-positive} in below.

\begin{lemma}
\label{lem:derivative}\[
y^{\prime}\left(s\right):=\frac{dy\left(s\right)}{ds}=f\left(t,s\right)+\int_{t}^{\infty}f\left(\eta,s\right)d\mu_{\eta}.\]
 
\end{lemma}

This lemma is proved by elementary calculation.

\begin{lemma}
\label{lem:unique-maxium-general}If $y^{\prime}\left(s\right)=0$
implies $y^{\prime\prime}\left(s\right)<0$ on $\left(0,1\right)$,
then one of the following three cases must be true:

1) $y^{\prime}\left(x\right)>0$ on $\left(0,1\right)$ and $\underset{s\in\left(0,1\right)}{\sup}y\left(s\right)=\underset{s\rightarrow1}{\lim}y\left(s\right)$; 

2) $f^{\prime}\left(x\right)<0$ on $\left(0,1\right)$ and $\underset{s\in\left(0,1\right)}{\sup}y\left(s\right)=\underset{s\rightarrow1}{\lim}y\left(s\right)$;

3) there exists a unique $s^{*}\in\left(0,1\right)$ such that $y^{\prime}\left(s^{*}\right)=0$,
and $\underset{s\in\left(0,1\right)}{\sup}y\left(s\right)=y\left(s^{*}\right)$.
\end{lemma}
\begin{proof}
Since the first two cases are trivial, we only prove the third case.
We shall prove that there exists a unique $s^{*}\in\left(0,1\right)$
s.t. $y^{\prime}\left(s^{*}\right)=0$. The existence is clear since
we have excluded the first two cases. The uniqueness is proved by
constructing a contradiction. Suppose that there are $z_{i}\in\left(0,1\right)$'s
s.t. $y^{\prime}\left(z_{i}\right)=0$. Take the largest $z_{l}<s^{*}$.
Since $y^{\prime\prime}\left(z_{l}\right)<0$ and $y^{\prime\prime}\left(s^{*}\right)<0$,
there exists a $\delta<\frac{x^{*}-z_{l}}{2}$ s.t. $y^{\prime}\left(z\right)<0$
on $\left(z_{l},z_{l}+\delta\right)$ and $y^{\prime}\left(z\right)>0$
on $\left(s^{*}-\delta,s^{*}\right)$. But this implies that there
exists $z^{\prime}\in\left[z_{l}+\delta,s^{*}-\delta\right]$ s.t.
$y^{\prime}\left(z^{\prime}\right)=0$, which contradicts the assumption
that $z_{l}<s^{*}$ is the largest root of $y^{\prime}\left(z\right)$.
\end{proof}

\begin{lemma}
\label{lem:formula-1og-positive}\[
\frac{2x}{1+x}+\log^{2}\left(1+x\right)-2\log\left(1+x\right)>0\]
 for all $x>0$.
\end{lemma}
\begin{proof}
Let \[
g\left(x\right)=\frac{2x}{1+x}+\log^{2}\left(1+x\right)-2\log\left(1+x\right).\]
 Since $g\left(0\right)=0$, this lemma is true if $g^{\prime}\left(x\right)>0$
for $x>0$. Note that \[
g^{\prime}\left(x\right)=\frac{2}{1+x}\left(\log\left(1+x\right)-\frac{x}{1+x}\right).\]
 We have $g^{\prime}\left(x\right)>0$ on $x>0$ if \[
\tilde{g}\left(x\right)=\log\left(1+x\right)-\frac{x}{1+x}>0\]
 on $x>0$. Since $\tilde{g}\left(0\right)=0$ and $\tilde{g}^{\prime}\left(x\right)=\frac{x}{\left(1+x\right)^{2}}>0$
on $x>0$, $\tilde{g}\left(x\right)>0$ on $x>0$. This lemma is proved. 
\end{proof}

In order to prove Theorem \ref{thm:optimal-s}, as the first step,
we  show that there exists $s^{*}\in\left(0,1\right)$ s.t. $y^{\prime}\left(s^{*}\right)=0$.
Note that \[
y^{\prime}\left(s\right)=\log\left(1+t\frac{1-s}{s}\right)-\int_{t}^{\infty}\frac{1}{1+\frac{1-\eta}{\eta}s}\cdot\frac{d\mu_{\eta}}{s}.\]
 It is easy to verify that $\lim_{s\rightarrow1}y^{\prime}\left(s\right)<0$.
Now let $s\rightarrow0$. Since \[
1+\frac{1-\eta}{\eta}s\ge1-s,\]
 \[
\int_{t}^{\infty}\frac{1}{1+\frac{1-\eta}{\eta}s}\frac{d\mu_{\eta}}{s}\le2\;\mathrm{as}\; s\le\frac{1}{2}.\]
 But \[
\underset{s\rightarrow0}{\lim}\log\left(1+t\frac{1-s}{s}\right)=\infty.\]
Then $\lim_{s\rightarrow0}y^{\prime}\left(x\right)>0$. We conclude
that $y^{\prime}\left(s^{*}\right)=0$ happens for some $s^{*}\in\left(0,1\right)$.

According to Lemma \ref{lem:unique-maxium-general}, it is sufficient
to prove that as $s\in\left(0,1\right)$, $y^{\prime}=0$ implies
$y^{\prime\prime}=0$. Set $y^{\prime}=0$. Then \[
\log\left(1+t\frac{1-s}{s}\right)-1+\int_{t}^{\infty}\frac{1-\eta}{1+\eta\frac{1-s}{s}}\cdot\frac{d\mu_{\eta}}{s}=0.\]
Now we calculate $y^{\prime\prime}$. \begin{align*}
y^{\prime\prime} & =\left(f\left(t,s\right)+\int_{t}^{\infty}f^{\prime}\left(\eta,s\right)d\mu_{\eta}\right)^{\prime}\\
 & =2f^{\prime}\left(t,s\right)+\int_{t}^{\infty}f^{\prime\prime}\left(\eta,s\right)d\mu_{\eta}\\
 & =-\frac{1}{s}\frac{2t\frac{1}{s}}{1+t\frac{1-s}{s}}+\frac{1}{s}\left(1-\int_{t}^{\infty}\left(\frac{1-\eta}{1+\eta\frac{1-s}{s}}\right)^{2}\frac{d\mu_{\eta}}{s}\right)\\
 & \overset{\left(a\right)}{\le}-\frac{1}{s}\left[\frac{2t\frac{1-s}{s}}{1+t\frac{1-s}{s}}-1+\left(\int_{t}^{\infty}\frac{1-\eta}{1+\eta\frac{1-s}{s}}\frac{d\mu_{\eta}}{s}\right)^{2}\right]\\
 & \overset{\left(b\right)}{=}-\frac{1}{s}\left[\frac{2t\frac{1-s}{s}}{1+t\frac{1-s}{s}}-1+\left(1-\log\left(1+t\frac{1-s}{s}\right)\right)^{2}\right],\end{align*}
where $\left(a\right)$ comes from the fact that $t\frac{1}{s}\ge t\frac{1-s}{s}$
and Jensen's inequality, and $\left(b\right)$ is from the assumption
$y^{\prime}=0$. Note that $t\frac{1-s}{s}>0$. By Lemma \ref{lem:formula-1og-positive},
$y^{\prime\prime}<0$. The $x^{*}\in\left(0,1\right)$ s.t. $y^{\prime}\left(s^{*}\right)=0$
is therefore unique and maximizes $y$. 

\bibliographystyle{IEEEtran}
\bibliography{Bib/_BC_Feedback,Bib/_Jindal,Bib/_Tse,Bib/_Dai,Bib/RandomMatrix,Bib/FeedbackMIMO_append}

\begin{biography}
[{\includegraphics[scale=0.07]{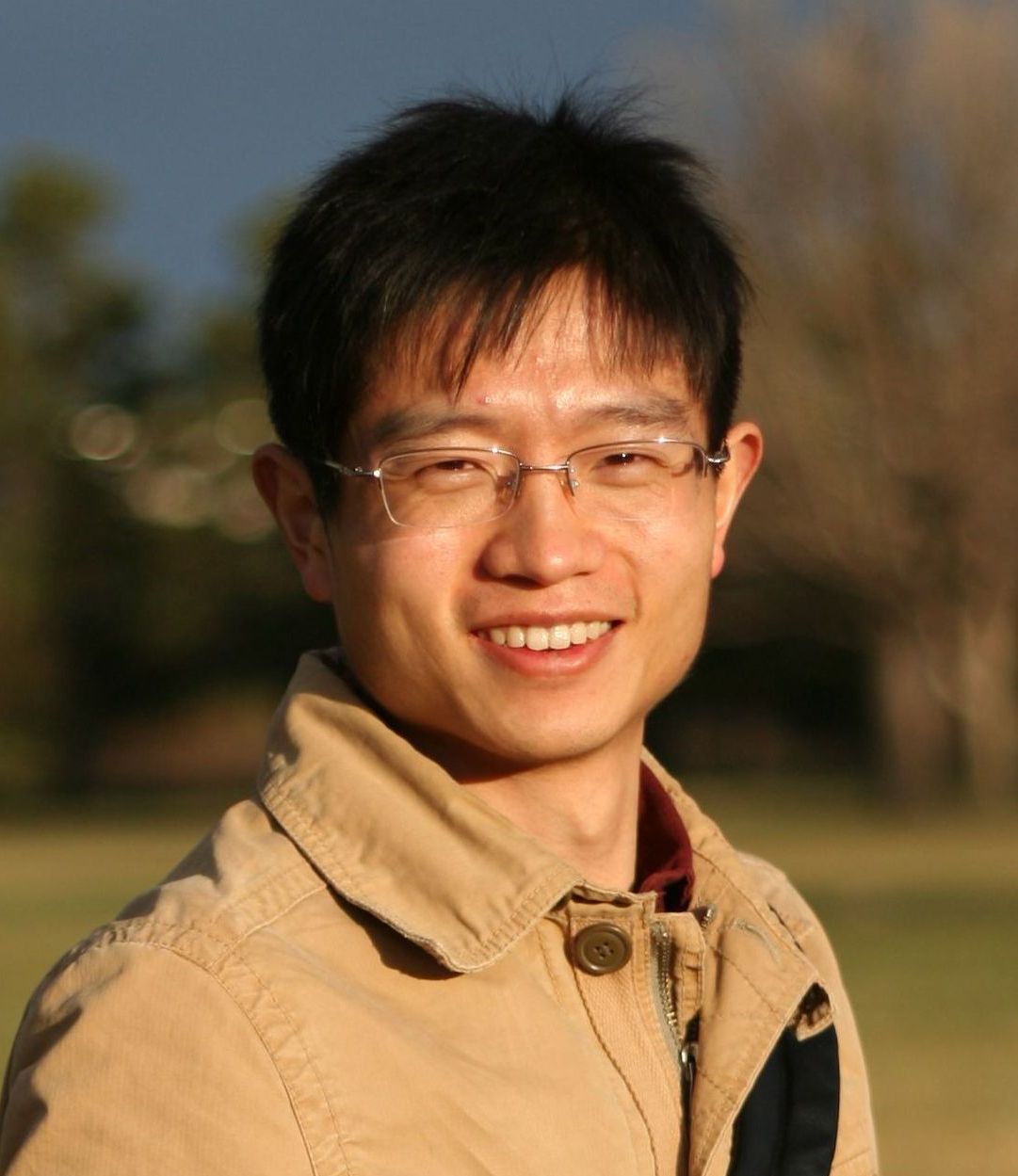}}]{Wei Dai} received
his Ph.D. and M.S. degree in Electrical and Computer Engineering from
the University of Colorado at Boulder in 2007 and 2004 respectively.
He is currently a Postdoctoral Research Associate at the Department
of Electrical and Computer Engineering, University of Illinois at
Urbana-Champaign. His research interests include communication theory,
information theory, compressive sensing, bioinformatics, and random
matrix theory.
\end{biography}

\begin{biography}
[{\includegraphics[scale=0.19]{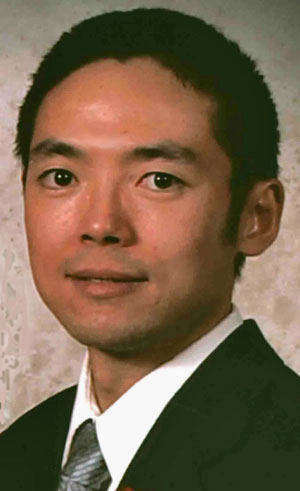}}]{Youjian (Eugene) Liu}
received the Ph.D. and M.S. degree in Electrical Engineering from
The Ohio State University in 2001 and 1998 respectively. Since August
2002, he has been an Assistant Professor with Department of Electrical
and Computer Engineering, University of Colorado at Boulder. From
January 2001 to August 2002, he worked on 3G mobile communication
systems as a Member of Technical Staff with Wireless Advanced Technology
Laboratory, Lucent Technologies, Bell Labs Innovations, Whippany,
New Jersey. His research interests include MIMO communications, coding
theory, and information theory. He is a recipient of 2005 Junior Faculty
Development Award at University of Colorado.
\end{biography}

\begin{biographynophoto}
{Brian Rider} received his Ph.D. in Mathematics from the Courant
Institute (New York University) in 2000. After a Lady Davis Fellowship
at the Technion, he had postdoctoral positions at Duke University
and MSRI. Since 2004 he has been an Assistant Professor of Mathematics
at the University of Colorado at Boulder. His research interests include
random matrix theory and spectral properties of random Schroedinger
operators. He is a recipient of a 2007 NSF CAREER grant as well as
a 2008 Rollo Davidson Prize.
\end{biographynophoto}

\begin{biography}
[{\includegraphics[scale=0.9]{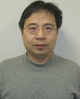}}]{Wen Gao} received
his Ph.D. and M.S. degree in Electrical Engineering from Purdue University
in 2001 and 1998 respectively. He joined Cooperate Research of Thomson
Inc. in Princeton in July 2001 working in the area of mobile and satellite
communications. His current research interests include cognitive radio,
multiple antenna technology and cooperative network, error control
coding and equalization. He also has extensive experiences in system
design and hardware development.
\end{biography}

\end{document}